\documentclass{article}

\usepackage{tabularx}
\usepackage{hyperref}
\usepackage{booktabs} 
\usepackage{float}
\usepackage[font=scriptsize]{caption}
\usepackage[font=scriptsize]{subcaption}
\usepackage{graphicx}
\usepackage{listings}
\usepackage{microtype}
\usepackage{verbatim}

\usepackage[
  paperheight=8.5in,
  paperwidth=5.5in,
  left=10mm,
  right=10mm,
  top=20mm,
  bottom=20mm]{geometry}
\usepackage[utf8]{inputenc}

\usepackage{graphicx}
\usepackage{wrapfig}
\usepackage[bottom]{footmisc}
\usepackage{listings}
\usepackage{enumitem}

\usepackage{wrapfig}
\usepackage{ragged2e}

\usepackage{array}
\usepackage[table]{xcolor}
\usepackage{multirow}
\usepackage{booktabs}
\usepackage{hhline}
\definecolor{palegreen}{rgb}{0.6,0.98,0.6}

\usepackage{amsmath}
\usepackage{amssymb}
\usepackage{multicol}
\usepackage{lipsum}
\usepackage{hyphenat}
\PassOptionsToPackage{hyphens}{url}
\usepackage{url}

\usepackage{rotating}

\usepackage[T1]{fontenc}
\usepackage{textcomp}
\usepackage{listings}
\usepackage{setspace}

\usepackage[numbers,sort]{natbib}

\newcommand*{\email}[1]{\small{\texttt{#1}}}

\renewcommand{\footnoterule}{%
  \kern -3pt
  \hrule width \textwidth height 0.5pt
  \kern 2pt
}

\date{}

\usepackage{titlesec}
\titleformat*{\section}{\large\bfseries}
\titleformat*{\subsection}{\normalsize\bfseries}
\titleformat*{\subsubsection}{\normalsize\bfseries}

\title{Evaluating the Limitations of Local LLMs in Solving Complex Programming Challenges \footnote{\protect
Preprint. Accepted to CCSC 2025. 
Copyright \copyright 2025 by the Consortium for Computing Sciences in Colleges.
Permission to copy without fee all or part of this material is granted provided
that the copies are not made or distributed for direct commercial advantage,
the CCSC copyright notice and the title of the publication and its date appear,
and notice is given that copying is by permission of the Consortium for
Computing Sciences in Colleges. To copy otherwise, or to republish, requires
a fee and/or specific permission.
The definitive version will be published by the Consortium for Computing Sciences in Colleges.

}}

\author{
Kadin Matotek, Heather Cassel, Md Amiruzzaman, Linh B. Ngo\\
Computer Science Department\\
West Chester University, West Chester, PA\\
\email{\{KM998744,HC946859,mamiruzzaman,lngo\}}@wcupa.edu
}

\begin{document}
\maketitle

\begin{abstract}

\noindent This study examines the performance of today’s open-source, locally hosted large-language models (LLMs) in handling complex competitive programming tasks with extended problem descriptions and contexts. Building on the original Framework for
AI-driven Code Generation Evaluation (FACE), the authors retrofit the pipeline to work entirely offline through the Ollama runtime, collapsing FACE’s sprawling per-problem directory tree into a handful of consolidated JSON files, and adding robust checkpointing so multi-day runs can resume after failures. The enhanced framework generates, submits, and records solutions for the full Kattis corpus of 3,589 problems across eight code-oriented models ranging from 6.7 billion to 9 billion parameters. The submission results show that the overall pass@1 accuracy is modest for the local models, with the best models performing at approximately half the acceptance rate of the proprietary models, Gemini 1.5 and ChatGPT-4. These findings expose a persistent gap between private, cost-controlled LLM deployments and state-of-the-art proprietary services, yet also highlight the rapid progress of open models and the practical benefits of an evaluation workflow that organizations can replicate on in-house hardware.


\end{abstract}


\section{Introduction}

Advances in Artificial Intelligence (AI) have led to the creation of some highly sophisticated large language models (LLMs). An LLM is an AI-based model that generates human-like text to perform various language-related tasks. Some of them, including the family of ChatGPT models and Gemini, are known to excel in code generation. These models have demonstrated the potential to automate certain software development tasks \cite{antoniades2024swe,lin2024llm, manish2024autonomous,
sridhara2023chatgpt}. However, their cloud-based and proprietary nature raises concerns such as data privacy, latency, and cost. Token quotas and rate limits hinder large-scale experiments, and usage fees can skyrocket when benchmarking thousands of problems. To avoid these issues, organizations now deploy open-source models locally. Yet, evaluations of their performance on complex coding tasks remain scarce.

Cloud-based models have been extensively benchmarked using competitive programming platforms \cite{dunder2024kattis, ngo2024face, nikolaidis2023end, yildiz2024assessing} and real-world tasks \cite{brown2020language, shen2023Advances}. The performance of local models has not yet been rigorously evaluated in similar contexts. In cases where local models have been evaluated \cite{coignion2024performance, souza2025code, tuttle2024can}, problem datasets typically lack quantity and variety. As a result, it becomes difficult to assess the models’ full problem-solving potential. This study evaluates several locally hosted LLMs through the Ollama platform \cite{ollama} on Kattis’ extensive collection of programming challenges \cite{kattis}.
Kattis is a publicly available platform with more than 3,500 coding problems of varying difficulty and is widely used to evaluate programming proficiency.

In this study, we build on the work of \cite{ngo2024face} by extending the Framework for AI-driven Code Generation Evaluation (FACE) and applying it to the same comprehensive set of Kattis problems. 
Whereas \cite{ho2024predicting} evaluated proprietary ChatGPT-4 and Gemini 1.5, we focus exclusively on locally hosted LLMs. By comparing our results with previous work, we aim to emphasize both the improvements and limitations of local LLMs in complex code generation tasks. This comparative analysis informs future research and industry practice about the trade-offs between local and cloud-based LLM solutions. To the best of our knowledge, no research has evaluated the performance of code generated by such a large number of open-source LLMs on a high volume of complex coding problems.


The paper is structured as follows. Section \ref{sec:review} reviews prior work. Section \ref{sec:framwork-extension} details our FACE framework extension. Section \ref{sec:experimentation} describes the experimental setup. Section \ref{sec:casestudy} presents the results. Section \ref{sec:conclusion} concludes and outlines future directions.

\section{Related Work}
\label{sec:review}
Several studies have examined both local and proprietary LLMs on code-related tasks: performance on coding platforms \cite{ngo2024face, yildiz2024assessing, coignion2024performance, ho2024predicting, manik2025chatgpt}, security of their suggestions \cite{elgedawy2024ocassionally, pearce2025asleep, perry2023users}, and bugs in generated code \cite{jesse2023large}. For example, Coignion et al. \cite{coignion2024performance} conducted an extensive evaluation of 18 code-oriented LLMs using a dataset of 204 Leetcode problems. Their experiments generated over 210,120 solutions—with GitHub Copilot contributing 2,040 solutions and eight open-source models yielding 12,240 each. Of the top three models, StarCoder achieved a pass@1 of 0.095 and a pass@10 of 0.132. CodeLlama scored 0.093 on pass@1 and 0.201 on pass@10, and GitHub Copilot followed closely with scores of 0.092 and 0.196, respectively. The pass@$k$ metrics represent the probability that at least one of $k$ solution attempts will succeed. The results indicate that open-source LLMs are capable of competitive performance. In some cases, their performance is comparable to industry-leading solutions such as Copilot when solving programming problems.


One of the limitations of these studies is that their problem sets often come from popular coding challenge sources such as LeetCode or Codeforces. This means that the problems' descriptions and corresponding solutions are well-known and often publicly discussed. On LeetCode, every problem includes a discussion tab, and solutions are often mirrored to GitHub and analyzed in blog posts.
Codeforces publishes official solutions within hours or days, and these are widely shared outside the site. In contrast, Kattis prohibits sharing solutions and routinely checks for plagiarism. As a result, solutions are scarce online,  making them underrepresented in training data \cite{sun2023head}. Large LLMs excel at questions based on widely available knowledge, but struggle with long-tail information \cite{kandpal2023large}. To our knowledge, only \cite{ho2024predicting, tran2023exploring} have evaluated online LLMs such as Gemini and ChatGPT using a small subset of Kattis problems and their built-in evaluation system.

Ngo et al. \cite{ngo2024face} introduce FACE (Framework for AI-driven Coding Generation Evaluation), a three-stage pipeline (Miner, Generator, Submitter) that automates end-to-end benchmarking of code-generation models on Kattis. In this work, we leverage and extend FACE to utilize local LLMs to generate solutions for more than 3,500 Kattis problems. This allows us to address the limitation noted above and extensively study how effective local LLMs are at solving complex competitive programming challenges.

\section{Framework Extension}
\label{sec:framwork-extension}

In extending the original FACE architecture \cite{ngo2024face} for our study, we improved the system's data organization, added support of locally hosted LLMs, and altered the submission process to Kattis for a more scalable and fault-tolerant approach. This section outlines how our modifications enable the evaluation of over 3,500 coding problems.

\subsection{FACE Architecture}

\begin{figure}[!h]
\centering
  \includegraphics[scale=0.3]{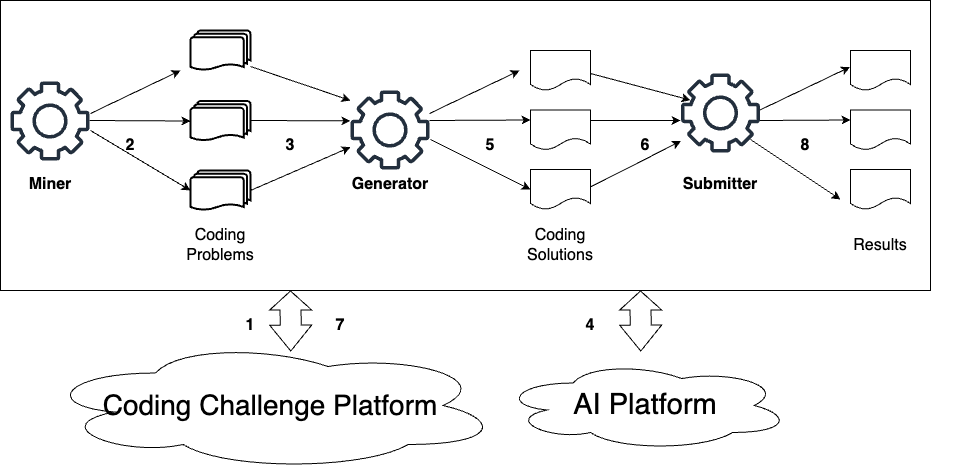}  
  \caption{Original FACE architecture, with numbers indicating each step in the process.}
  \vspace{-0.3in}  
\label{FACE_framework}
\end{figure}

FACE \cite{ngo2024face} automates collecting thousands of programming problems from a coding-challenge platform, querying an AI service for solutions, and submitting those solutions back for evaluation. Figure~\ref{FACE_framework} represents the original FACE framework.


\vspace*{-2pt} 

\subsection{Change in Data Organization}
Table~\ref{tab:face_vs_json} summarizes both FACE’s original structure and our new JSON‐based implementation.
FACE’s original architecture stored each problem in its own directory, with subfolders for metadata, problem statements, test cases, and submission logs. Although straightforward, this doesn’t scale: processing more than 3,500 problems across just two models produced over 25,000 individual .txt/.py files plus a similar number of folders. To address this, we adopted a JSON-based design that consolidates all problem data, model outputs, and submission results into just 17 JSON files.
There is one file for the problem data and two files per model—one for outputs and one for the submission results. This reduced the file count by roughly 99.9\%, even as we scaled from two to eight models. Consolidating into a handful of JSON files simplifies maintainability and version control, improves performance, and increases portability.

\subsection{Integration of Locally Hosted Models}

Our extension departs from the original FACE framework by integrating locally run LLMs, rather than relying on cloud-based APIs. The enhanced ``Generator'' component was modified to accommodate locally hosted models via a standardized interface. In this phase, each model receives a standard JSON payload (\texttt{kattis\_problems.json}) containing problem descriptions and test cases extracted using FACE. The responses for each model were stored in separate JSON files.

\begin{table}[H]
\centering
\begin{tabular}{
  >{\raggedright\arraybackslash}p{0.45\textwidth}
  >{\raggedright\arraybackslash}p{0.45\textwidth}
}
\toprule
\textbf{Original FACE Structure} & \textbf{JSON–Based Implementation} \\
\midrule
One directory per problem:
\begin{itemize}\scriptsize
  \item metadata file
  \item problem text (.txt)
  \item test cases (.txt)
  \item results (logs)
  \item submissions (code)
\end{itemize}
&
Two JSON files per model:
\begin{itemize}\scriptsize
  \item \verb|kattis_problems.json|: all metadata, statements, and test cases
  \item \verb|solutions_<model name>.json|: full responses + code
  \item \verb|submissions_<model name>.json|: solution statuses from Kattis
\end{itemize}
\\
\bottomrule
\end{tabular}
\caption{Comparison of the original FACE directory structure with our consolidated JSON-based implementation.}
\label{tab:face_vs_json}
\end{table}

\subsection{Improvements in the Submission Process}

To transition from solution generation to evaluation, we extended the ``Submitter'' component of the FACE architecture. This new submission process emphasizes consistency and recoverability. Key aspects of our enhancements include:

\begin{itemize}[
  leftmargin=1.5em,
  itemsep=0.8em,    
  parsep=0pt,       
  topsep=1pt        
]
  \item \textbf{\small Atomic Submission and Results Storage:}  
    For each problem, the generated Python solution is written to a temporary file and submitted via the official Kattis API. The system uses atomic file operations: The temporary results file is flushed and synced to disk before being renamed to its final JSON output. This careful writing procedure minimizes data loss by ensuring that each submission’s result is reliably stored.
    
  \item \textbf{\small Checkpointing:}  
    Because submissions for each model took several days to complete, there was a risk of data corruption during the process. To address this overhead, we implemented a checkpoint mechanism. A designated checkpoint file maintains the identifier of the last successfully processed problem. Upon interruption, the process reads the checkpoint to resume from where it left off. This prevents reprocessing problems that have already been evaluated. This checkpointing strategy greatly enhances the framework's resilience to any disruptions.
\end{itemize}

\section{Experimentation}
\label{sec:experimentation}

In this section, we detail the experimental setup used to benchmark eight locally hosted LLMs on more than 3,500 Kattis problems using our extended FACE pipeline. We begin by describing the selected models and summarizing their training characteristics. Next, we outline the computing environment and hardware configuration. Finally, we explain how the solutions were generated, submitted, and collected for evaluation.

\subsection {Model Descriptions}

We selected models with 6.7–9 billion parameters so that they could run on small-scale servers. Although we conducted experiments on a departmental GPU server, we also tested on the authors’ machines with consumer-grade GPUs (8 GB VRAM) and observed only negligible slowdowns in solution generation. Our choices were based on models available on the Ollama platform.
Table~\ref{tab:models_summary} summarizes the eight local LLMs chosen in our study.

\setlength{\tabcolsep}{4pt}
\begin{table}[H]
\centering
\small
\begin{tabularx}{\linewidth}{X c c c c}
\toprule
\textbf{Model} & \textbf{Params (B)} & \textbf{Size (GB)} & \textbf{Pre-training} & \textbf{Context (K)} \\
\midrule
CodeLlama \cite{rozière2024codellamaopenfoundation}         & 7.0  & 3.8 & 500 B  & 16    \\
CodeQwen  \cite{bai2023qwentechnicalreport}                  & 7.0  & 4.2 & 90 B   & 64    \\
DeepSeek-Coder  \cite{guo2024deepseekcoderlargelanguagemodel} & 6.7  & 3.8 & 2 T    & 16    \\
DolphinCoder \cite{dolphinCoder}                            & 7.0  & 4.2 & —      & 16    \\
Granite-Code  \cite{mishra2024granitecodemodelsfamily}       & 8.0  & 4.6 & 4.5 T  & 125   \\
Llama 3.1  \cite{grattafiori2024llama3herdmodels}            & 8.0  & 4.9 & 15.6 T & 128   \\
Qwen2.5-Coder  \cite{hui2024qwen25codertechnicalreport}      & 7.0  & 4.7 & 5.2 T  & 32    \\
Yi-Coder  \cite{ai2025yiopenfoundationmodels}                & 9.0  & 5.0 & 3.1 T  & 128  \\
\bottomrule
\end{tabularx}
\caption{Summary of local LLMs: parameter count (in billions of parameters), model size (in gigabytes), total pre-training (in tokens), and context window size (in thousands of tokens).}
\label{tab:models_summary}
\end{table}

To understand their design choices, it is helpful to know which models are built on existing foundations versus those trained entirely from scratch. CodeLlama was initialized with Llama 2 weights and fine-tuned on 500 B code tokens using an infilling objective. Llama 3.1  belongs to the newer Llama 3 family, pre-trained on 15.6 T multilingual tokens and further aligned through supervised fine-tuning and direct preference optimization. CodeQwen continues pre-training of the original Qwen base model on 90 B mixed text–code tokens, optimized for code generation with Flash Attention \cite{dao2022fastattention}. Qwen2.5-Coder builds on the Qwen 2.5 architecture and is further pre-trained on a 5.2 T token mix of 70\% code, 20\% text–code grounding data, and 10\% math data. DolphinCoder extends StarCoder2 \cite{lozhkov2024starcoder2stackv2} and is trained on the LeetCode Rosetta dataset \cite{hartford2023leetcode} for competitive programming. DeepSeek-Coder and Granite-Code are trained from scratch: DeepSeek-Coder on 2 T tokens (87\% code, 10\% English code text, 3\% Chinese) and Granite-Code on 4.5 T tokens (8\% code, 20\% natural language). Finally, Yi-Coder was trained from scratch on a 3.1 T token bilingual (English/Chinese) web crawl with extensive filtering and deduplication.


\subsection{Running the Experiments}
\label{sec:computing}

The experiment was carried out on the department’s GPU server, which runs Ubuntu 22.04 LTS. The server consists of dual Intel Xeon 6426Y CPUs (64 cores total; 56 available), dual NVIDIA L4 GPUs, each with 24 GB of memory, 256 GB of RAM, and 23 TB of SSD storage. To interact with the LLMs, we used Ollama version 0.3.4 and CUDA version 12.8.

In the first phase of the experiments, the modified FACE platform contacted an external Ollama server and instructed it to load a specific LLM. Next, it used the loaded model to generate solutions for all Kattis problems. Lastly, FACE gradually submitted them to Kattis and collected the results. The total runtime for these experiments exceeded three weeks.


\section{Performance Evaluation}
\label{sec:casestudy}

This section presents a detailed evaluation of model performance across three dimensions: solution generation speed, correctness (\textit{Accepted} status), and failure types (e.g., \textit{Wrong Answer}, \textit{Run Time Error}). We analyzed results across problem difficulties (Easy, Medium, and Hard) to uncover trends and limitations of locally hosted LLMs.

\subsection {Generation Time Comparison}
\label{sec:generation-time}

To assess efficiency, we measured response generation times for all 3,589 problems. Table \ref{tab:runtime} presents summary statistics, Figure \ref{fig:histograms} shows generation time histograms per model, and Figure \ref{cdf} shows the corresponding cumulative distribution functions (CDF).

\begin{table}[htb!]
\centering
\caption{Response generation time (in seconds) for each model on the full dataset, ordered by parameter size. Models with the two highest acceptance rates are shown in bold.}
\setlength{\tabcolsep}{4pt} 
\small    
\begin{tabular}{lccccc}
\toprule
\textbf{Model Name}   & \textbf{Mean (s)} & \textbf{Median (s)} & \textbf{Std (s)} & \textbf{Min (s)} & \textbf{Max (s)} \\
\midrule
DeepSeek-Coder        & 15.71 & 12.89 & 62.48  &  1.38 & 2163.78 \\
CodeLlama             & 16.11 & 10.84 & 77.94  &  0.82 & 1815.62 \\
CodeQwen              & 10.48 &  8.38 & 50.09  &  0.70 & 1738.14 \\
DolphinCoder          & 20.02 & 10.59 &117.78  &  0.59 & 1775.38 \\
\textbf{Qwen2.5-Coder}         & 14.82 & 14.00 & 31.70  &  2.24 & 1890.25 \\
Granite-Code          &  8.35 &  7.39 &  6.21  &  0.26 &  216.92 \\
Llama3.1              & 12.53 & 11.64 &  4.88  &  3.08 &   58.62 \\
\textbf{Yi-Coder}              & 14.64 & 12.72 & 50.59  &  1.41 & 2149.67 \\
\bottomrule
\end{tabular}
\label{tab:runtime}
\end{table}

\subsection{Outliers}

\begin{figure}[!h]
  \centering
  \includegraphics[scale=0.25]{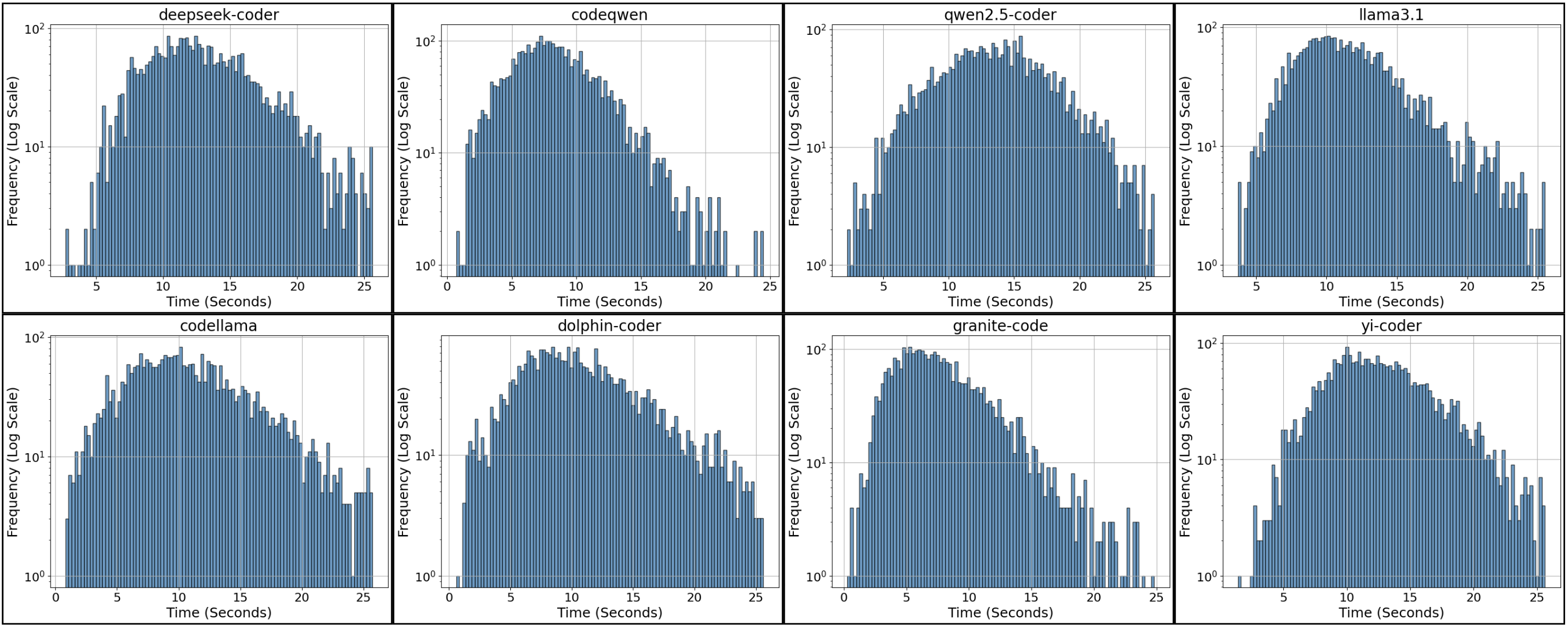}
  \vspace{-0.1in}
  \caption{Histograms of generation times (log‐scale frequency) for each model.}
  \label{fig:histograms}
\end{figure}

\begin{figure}[!h]
\centering
  \includegraphics[scale=0.30]{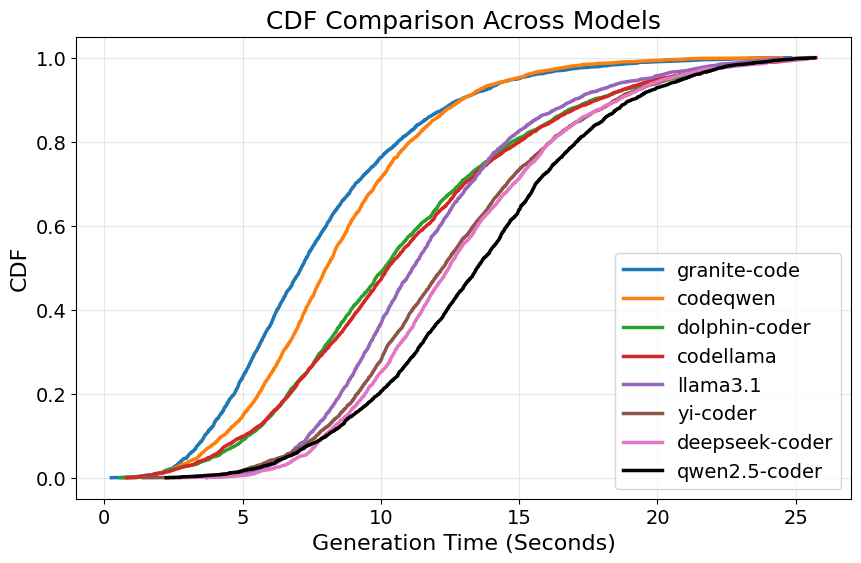}  
  \vspace{-0.1in}  
  \caption{CDF comparison of generation time across models.} 
\label{cdf}
\end{figure}

A small fraction of model–problem pairs exhibited exceptionally large generation times (see Table~\ref{tab:runtime}). To prevent these rare spikes from dominating our future histograms and CDFs (Figs.~\ref{fig:histograms} and \ref{cdf}), we applied the conventional box plot rule and discarded any outliers.
\\
\\
For each model’s response times, we computed the first and third quartiles ($Q_1$, $Q_3$), to form the interquartile range:
\[
\mathrm{IQR} = Q_3 - Q_1,
\]
and labeled any value outside the range given by the 1.5 IQR Rule:
\[
\bigl[\,Q_1 - 1.5\,\mathrm{IQR},\;Q_3 + 1.5\,\mathrm{IQR}\bigr]
\]
as an outlier.  The number of excluded outliers per model is reported in Table~\ref{tab:outlier_counts}.

\begin{table}[H]
\centering
\caption{Number of generation-time outliers
excluded per model, sorted from highest to lowest. Models with the two highest acceptance rates are shown in bold}
\small
\setlength{\tabcolsep}{4pt}
\begin{tabular}{lc}
\toprule
\textbf{Model Name}          & \textbf{\# Outliers} \\
\midrule
Llama3.1                  & 149 \\
DolphinCoder              & 147 \\
CodeLlama                 & 127 \\
Granite-Code              & 117 \\
DeepSeek-Coder            & 114 \\
CodeQwen                  & 113 \\
\textbf{Yi-Coder}         &  76 \\
\textbf{Qwen2.5-Coder}    &  65 \\

\bottomrule
\end{tabular}
\label{tab:outlier_counts}
\end{table}

Although Llama3.1 has the lowest maximum generation time (58.62 s; Table~\ref{tab:runtime}), it exhibits the most outliers (149; Table~\ref{tab:outlier_counts}). This occurs because its response times are tightly clustered, as shown by a low standard deviation. This yields a small interquartile range. As a result, even modest spikes exceed the outlier threshold. In contrast, Yi-Coder and Qwen2.5-Coder—the two best-performing models by acceptance rate—had the fewest outliers, 76 and 65, respectively.
    
\vspace*{-2pt} 

\vspace*{-3pt} 

Figure \ref{cdf} illustrates the Cumulative Distribution Function (CDF) for the generation times of different models. Each curve represents the CDF of a specific model, showing the proportion of solutions generated within a given time threshold. The x-axis corresponds to the generation time (in seconds), while the y-axis represents the CDF, ranging from 0–1, where 1 indicates that all solutions were generated within that time.

From the plot, it is evident that models such as Qwen2.5-Coder, DeepSeek-Coder, and Yi-Coder exhibited the slowest solution generation times, with their CDF curves increasing more gradually and extending further along the x-axis. Despite their slower speeds, these models performed the best, with average acceptance rates of 5.7\%, 2.9\%, and 5.4\%, respectively, as shown in Table \ref{tab:ac_rate_comparison}. This comparison underscores the trade-off between computational efficiency and generation speed in these models.

\subsection{Correctness and Failure Analysis}

First, we assessed how each model's submissions performed in terms of correctness and common failure modes. For every problem difficulty level (Easy, Medium, and Hard), we counted the number of submissions that resulted in an accepted solution, as well as those that failed due to compilation errors, wrong answers, run‐time errors, time‐out, or time/memory limits. This breakdown helped us identify not only which models could solve a given fraction of problems, but also the types of mistakes they most frequently made. Tables~\ref{tab:easy_counts}–\ref{tab:hard_counts} summarize these counts for each model across the three difficulty tiers.

\begin{table}[H]
\scriptsize
\centering
\caption{Outcome counts per difficulty by model (Easy), ordered by model size. The two top-performing models are shown in bold.}
\setlength{\tabcolsep}{2pt} 
\resizebox{\linewidth}{!}{   
\begin{tabular}{@{}lcccccccc@{}}
\hline\hline
\multicolumn{9}{c}{\textbf{Problem Difficulty: Easy}} \\
\hline
\textbf{Status} & deepseek & codellama & codeqwen & dolphin & \textbf{qwen2.5} & granite & llama3.1 & \textbf{yi-coder} \\
\hline
\textit{Accepted} &   90 &   16 &   49 &   21 &  157 &   14 &   91 &  139 \\
\textit{Compile Error}  &   27 &  210 &   21 &   92 &    2 &   28 &    8 &   20 \\
\textit{Wrong Answer}  &  381 &  245 &  326 &  370 &  323 &  485 &  318 &  368 \\
\textit{Run Time Error}  &  114 &  145 &  209 &  127 &  120 &   82 &  185 &   82 \\
\textit{Time Limit Exceeded} &    8 &    4 &    8 &    6 &   17 &    2 &   18 &   10 \\
\textit{Memory Limit Exceeded} &    0 &    0 &    0 &    0 &    1 &    0 &    0 &    1 \\
\textbf{Total} &  620 &  620 &  613 &  616 &  620 &  611 &  620 &  620 \\
\hline\hline
\end{tabular}
}
\label{tab:easy_counts}
\end{table}

\begin{table}[H]
\scriptsize
\centering
\caption{Outcome counts per difficulty by model (Medium), ordered by model size. The two top-performing models are shown in bold.}
\setlength{\tabcolsep}{2pt} 
\resizebox{\linewidth}{!}{
\begin{tabular}{@{}lcccccccc@{}}
\hline\hline
\multicolumn{9}{c}{\textbf{Problem Difficulty: Medium}} \\
\hline
\textbf{Status} & deepseek & codellama & codeqwen & dolphin & \textbf{qwen2.5} & granite & llama3.1 & \textbf{yi-coder} \\
\hline
\textit{Accepted}   &   15 &    1 &    5 &    1 &   47 &    4 &    8 &   52 \\
\textit{Compile Error}  &   72 &  392 &   59 &  216 &   12 &   70 &   17 &   37 \\
\textit{Wrong Answer}    &  770 &  506 &  543 &  717 &  719 &  993 &  645 &  819 \\
\textit{Run Time Error}   &  375 &  362 &  608 &  311 &  422 &  184 &  549 &  300 \\
\textit{Time Limit Exceeded}  &   38 &   17 &   48 &   25 &   74 &   10 &   55 &   58 \\
\textit{Memory Limit Exceeded} &    8 &    0 &    4 &    1 &    4 &    2 &    4 &   12 \\
\textbf{Total}  & 1278 & 1278 & 1267 & 1271 & 1278 & 1263 & 1278 & 1278 \\
\hline\hline
\end{tabular}
}
\label{tab:medium_counts}
\end{table}

\begin{table}[H]
\scriptsize
\centering
\caption{Outcome counts per difficulty by model (Hard), ordered by model size.}
\setlength{\tabcolsep}{2pt} 
\resizebox{\linewidth}{!}{
\begin{tabular}{@{}lcccccccc@{}}
\hline\hline
\multicolumn{9}{c}{\textbf{Problem Difficulty: Hard}} \\
\hline
\textbf{Status}  & deepseek & codellama & codeqwen & dolphin & qwen2.5 & granite & llama3.1 & yi-coder \\
\hline
\textit{Accepted}  &     0 &     0 &     0 &     0 &     0 &     0 &     1 &     1 \\
\textit{Compile Error}   &   128 &   475 &    95 &   291 &    18 &    85 &    35 &    80 \\
\textit{Wrong Answer}  &   985 &   719 &   646 &   853 &   924 &  1279 &   833 &  1096 \\
\textit{Run Time Error}  &   512 &   478 &   870 &   511 &   615 &   290 &   773 &   418 \\
\textit{Time Limit Exceeded}   &    48 &    16 &    60 &    28 &   120 &    15 &    43 &    75 \\
\textit{Memory Limit Exceeded} &    11 &     0 &     6 &     1 &    12 &     2 &     5 &    18 \\
\textbf{Total} &  1684 &  1688 &  1677 &  1684 &  1689 &  1671 &  1690 &  1688 \\
\hline\hline
\end{tabular}%
}
\label{tab:hard_counts}
\end{table}

\noindent\textit{Note:} Some difficulty totals differ from the full problem set size because we filtered out any statuses outside the six tracked categories (\textit{Accepted}, \textit{Compile Error}, \textit{Memory Limit Exceeded}, \textit{Run Time Error}, \textit{Time Limit Exceeded}, \textit{Wrong Answer}).

For \textbf{Easy problems}, all models performed modestly, with Qwen2.5-coder and Yi-Coder obtaining the highest number of accepted submissions, with 157 and 139, respectively. However, even at this level, the majority of attempts resulted in \textit{Wrong Answers} or \textit{Run Time Errors}, suggesting difficulty in reliably solving even simpler problems.

As difficulty increased, performance deteriorated significantly. For \textbf{Medium problems}, most models achieved fewer than 10 accepted solutions; only Yi-Coder and Qwen2.5-coder showed slight improvements, 52 and 47, respectively.
\textit{Wrong Answer} and \textit{Run Time Error} rates were consistently high across all models, indicating challenges in handling intermediate problem complexity.

On \textbf{Hard problems}, no models could reliably generate correct solutions. Only Yi-Coder and Llama3.1 managed to produce even a single accepted response. This result underscores a significant performance ceiling for current local LLMs in high-difficulty settings.

\subsection{Comparison with Cloud-Based Models}
\label{sec:cloud_comparison}

Before presenting the acceptance rates of our local models, we recap the cloud-based benchmarks reported by Ngo et al. \cite{ngo2024face}. In their evaluation of 1,981 Kattis problems, Gemini 1.5 achieved an acceptance rate of 10.9\% (217/1,981) and ChatGPT-4 achieved 10.7\% (211/1,981). When these results are broken down by problem difficulty (on a 1.0–10.0 scale), they found that among the 95 problems rated 1.0–2.0, 68\% (65/95) were accepted. This fell to 21\% (48/225) in the 2.0–4.0 range, and acceptance was effectively zero beyond the 4.0 rating. Their findings illustrate how performance drops off rapidly as problem difficulty increases.

\begin{table}[!h]
\scriptsize
\centering
\caption{Acceptance rates for cloud-based and local models, sorted from highest to lowest.}
\begin{tabular}{lc}
\toprule
\textbf{Model}               & \textbf{AC Rate (\%)} \\
\midrule
\multicolumn{2}{l}{\textit{Cloud-based ($n = 1{,}981$)}} \\
\\
Gemini 1.5                       & 10.9                  \\
ChatGPT-4                      & 10.7                  \\
\midrule
\multicolumn{2}{l}{\textit{Local models ($n = 3,589$)}} \\
\\
Qwen2.5-Coder                &  5.7                  \\
Yi-Coder                     &  5.4                  \\
DeepSeek-Coder               &  2.9                  \\
Llama3.1                     &  2.8                  \\
CodeQwen                     &  1.5                  \\
DolphinCoder                 &  0.6                  \\
CodeLlama                    &  0.5                  \\
Granite-Code                 &  0.5                  \\

\bottomrule
\end{tabular}
\label{tab:ac_rate_comparison}
\end{table}

\noindent The acceptance rates reported in Table~\ref{tab:ac_rate_comparison} correspond to our \textbf{pass@1} metric, i.e., the probability that a single generated solution for a problem is correct. Formally, for any \(k \ge 1\),
\begin{equation}
  \label{eq:passmetrics}
  \mathrm{pass@}k \;=\;\mathbb{E}_{\mathrm{Problems}}\Biggl[1 \;-\;\frac{\binom{n - c}{k}}{\binom{n}{k}}\Biggr]
\end{equation}
where 
\begin{itemize}
 \item \(\mathbb{E}_{\mathrm{Problems}}\) denotes the expectation taken over the distribution of problems
  \item \(n\) is the total number of samples generated per problem
  \item \(c\) is the number of those samples that pass all tests 
  \item \(k\) is the number of samples considered
\end{itemize}
In the special case \(k=1\), equation~\eqref{eq:passmetrics} simplifies to
\[
  \mathrm{pass@1}
  \;=\;\mathbb{E}_{\mathrm{Problems}}\Bigl[\tfrac{c}{n}\Bigr],
\]
so Table~\ref{tab:ac_rate_comparison}’s acceptance rates measure the average fraction of problems for which the model’s first attempt succeeded.




\section{Conclusion}
\label{sec:conclusion}

In this study, we evaluated eight open-source LLMs against more than 3,500 programming problems. We extended the FACE framework to work with Ollama, consolidated all data into JSON files, and added a checkpointing system. This infrastructure enabled large-scale testing and direct comparison with previous cloud-based results. The local models with the highest performance, Qwen2.5- Coder and Yi-Coder, achieved pass@1 rates of 5.7\% and 5.4\%, approximately half the pass@1 rates of ChatGPT and Gemini. The trade-off here is that local models can be run an unlimited number of times compared with the token limitation or the monetary cost of the proprietary LLM solution. 

Our findings underscore both the promise and the current limitations of locally hosted LLMs for complex code generation. Going forward, we will look at fine-tuning these models on task-specific datasets, combining local and cloud-based inference in hybrid workflows, and using smarter prompt designs or built-in debugging steps to improve the correct submission rate. As the open-source model landscape continues to grow, benchmarks become increasingly vital. Large and thorough evaluations guide model improvements and inform deployment decisions. This is especially important in settings where data privacy or limited computing resources are key concerns.

Beyond the raw accuracy numbers, our findings carry direct weight for educators seeking to modernize assessment. Local code-centric LLMs can power explain-as-you-grade autograders that return step-wise diagnostics instead of the usual binary pass/fail, while keeping student data safely on campus hardware. Lightweight Ollama deployments lower the cost threshold for in-IDE tutoring agents that gently assist novice programmers without handing them the full solution. Instructors gain the freedom to iterate rubrics rapidly, run large-batch regrading overnight, and even spin up model variants tuned to specific curricula. Taken together, the results point toward a future where scalable private feedback loops complement traditional office hours rather than replace them. 

\vspace*{-2pt} 

\medskip

\bibliographystyle{abbrvnat}
\bibliography{sample}

\end{document}